\documentclass[final,5p,times,twocolumn]{elsarticle}

\usepackage{epsfig}
\usepackage{graphicx}
\usepackage{subcaption}
\usepackage{booktabs}
\usepackage{siunitx}

\usepackage{amssymb}
\usepackage{amsmath}
\usepackage{bm}
\usepackage{amsthm}

\usepackage{lineno}

\usepackage{xcolor}

\journal{TBD}

\begin{document}

\begin{frontmatter}



\title{Influence of Droplet Morphology on Fluid Dynamics and Impact Erosion of Wind Turbine Blades}

\author[a]{Hao Hao\corref{co}}
\cortext[co]{Corresponding author}
\ead{hh3117@ic.ac.uk}
\author[a]{Antonis Sergis}
\author[a]{Maria N. Charalambides}
\author[a]{Alex M. K. P. Taylor}
\author[a]{Yannis Hardalupas}

\affiliation[a]{organization={Department of Mechanical Engineering, Imperial College London},
            city={London},
            postcode={SW7 2AZ}, 
            country={United Kingdom}}

\begin{abstract}
The impact of droplets, such as rain and sea sprays, on wind turbine blade surfaces can cause severe damage through leading-edge erosion (LEE). Current industrial practice has largely relied on empirical fatigue models to predict damage progression and guide maintenance. However, an important limitation of existing models is the assumption on ideal impact conditions, including perfectly spherical droplet and normal incidence, which rarely occur in reality. This study develops a fluid-structure interaction (FSI) model that couples incompressible multiphase flow simulations with finite element (FE) analysis of blade materials. Using this framework, we investigate the impact dynamics of droplets with arbitrary shapes and orientations. The results reveal a positive correlation between the peak impact pressure, the influential area at the pressure peak, and the peak impact force. In contrast, the time of peak force scales inversely due to momentum conservation. From these findings, we propose a \textit{universal scaling law} that relates all key impact loadings to the simple parameter of the droplet \textit{projection area}, which can be determined solely from the initial droplet geometry and orientation. This law provides a predictive framework for impact loading dynamics of non-ideal shapes, offering new insights into the mechanisms of LEE and improving the accuracy of erosion predictions for wind turbine blades.
\end{abstract}



\begin{keyword}
Key words \sep Key words


\end{keyword}

\end{frontmatter}

\nolinenumbers


\section{Introduction}
\label{introduction}
Impacts between liquid droplets and solid surfaces frequently cause erosion problems in engineering applications, including aircraft surface erosion \cite{Gohardani2011}, steam turbine blade erosion \cite{Preece1980}, and wind turbine blade leading-edge erosion \cite{Bech2018}. Among these, leading-edge rain erosion has emerged as a particularly severe issue for modern wind turbines. Raindrops and sea sprays repeatedly strike the blade surfaces, producing material damage through a fatigue-driven mechanism \cite{Springer1976,Eisenberg2018}. On large-scale turbines, with blade lengths exceeding $100$ m \cite{Gert2012,Verma2020}, droplets can impact at velocities approaching $100$ \unit{\meter\per\second} near the blade tips. Such high-speed impingements can rapidly induce surface roughening \cite{David2020,Muthu2024} or even structural failure \cite{Mishnaevsky2020}. Consequently, wind turbine blades now require more frequent and effective maintenance than ever before \cite{Javier2024,Herring2019}, particularly in offshore environments where operating conditions are more severe \cite{Dao}.

From a fluid-dynamic perspective, the dimensionless parameters governing high-speed impact erosion, namely the Reynolds number $Re$ and Weber number $We$, are both much greater than unity:
\begin{align}
    Re&=\frac{\rho ul}{\mu} \\
    We&=\frac{\rho u^2l}{\sigma}
\end{align}
where $\rho$, $u$, $l$, $\mu$, and $\sigma$ denote the liquid density, impact velocity, characteristic length, dynamic viscosity, and surface tension, respectively. Moreover, since the impact velocity of water droplets is much lower than the speed of sound in water, the liquid can be regarded as incompressible. Consequently, droplet impact and spreading are mainly inertia-driven, with viscous and capillary forces playing only a negligible role \cite{Riboux2014}. This is particularly true during the initial phase of impact \cite{Egger2010, Sanjay2025}, when the loading on the surface is most severe \cite{Cheng2019}, and thus the stage of greatest relevance for impact erosion.

Analytical studies on droplet impact date back to the pioneering work of Worthington, who investigated the morphologies of droplets impacting a solid surface \cite{Worthington1876} or a solid surface covered by a liquid film \cite{Worthington1908}. With regard to impact loadings, von Kármán \cite{Karman} and Wagner \cite{Wagner} provided the first analytical solutions for impact force and pressure, respectively, in the context of seaplanes float landings. Wagner's pressure solution, derived within the framework of inviscid potential flow theory, has since been refined by Cointe and Armand \cite{Cointe1987}, Oliver \cite{Oliver202}, and more recently by Hao \textit{et al.} \cite{Hao_ana}. While these analytical models focused on the early impact stage under invscid scheme, Eggers and co-workers \cite{Egger2010,Lagubeau2012} noted the importance of viscous and cappillary effects, and proposed a time-dependent hyperbolic flow solution for the late-time spreading and deposition regimes, where impact loading decays.

The dependence of the maximum impact force on the dimensionless parameters $Re$ and $We$ has been examined experimentally by Cheng \textit{et al.} \cite{Cheng2019}, Zhang \textit{et al.} \cite{Zhang2022} and Sanjay and co-workers \cite{Sanjay2025, Sanjay2025_2}. In parallel, Roisman \textit{et al.} \cite{Roisman2009, Roisman2009_2}, Gordillo \textit{et al.} \cite{Gordillo2019, Garcia2020}, and Sanjay and Lohse \cite{Sanjay2025} investigated the dependence of late-stage behaviours, such as the maximum spreading radius, on $Re$ and $We$. Mitchell \textit{et al.} \cite{Mitchell} interpolated the inertia-driven impact regime to the spreading regime using an extensive experimental dataset, thereby obtaining an empirical model that captures the proper decay of the impact force. Other studies have addressed specific mechanisms: Thoroddsen \textit{et al.} \cite{Thoroddsen2005} and Garc{\'i}a-Geijo \textit{et al.} \cite{Garcia2024} investigated air bubble entrapment and cushioning at initial droplet impact; Zhang \textit{et al.} \cite{Zhang2022} and Sanjay \textit{et al.} \cite{Sanjay2025_2} reported the intriguing occurrence of a \textit{second peak} in the impact force as the droplet jumps off hydrophobic surfaces. For a comprehensive review of recent progresses in droplet impact studies, readers are referred to \cite{Josserand2016}.

Numerical studies on droplet impact erosion typically employ finite element (FE)-based methods \cite{Keegan2012}, including the coupled Eulerian-Lagrangian (CEL) method within a fully finite element (FE) framework, and a hybrid smoothed particle hydrodynamics (SPH)-FE method. However, FE-based methods often suffer from limited resolution of the liquid phase, or require prohibitively high computational resources, due to the large deformations involved in the liquid phase \cite{Zhou2009}. Alternative hybrid approaches thus have been implemented\textemdash for example, coupling an independent computational fluid dynamics (CFD) simulation of the liquid phase \cite{Amirzadeh2017, Nick2023, Hao_cav} or an analytical solution of droplet impact \cite{Hao_ANCM} with an FE analysis of the solid phase. These methods have demonstrated notable success in capturing both fluid dynamics and material response. Other advanced fluid-structure interaction (FSI) simulations in related fields\textemdash though not specifically on droplet impact erosion\textemdash include multiphase bubble-liquid-solid models \cite{Zeng2022, Reese2023, Heidary2024}. In most such numerical studies, the droplet flow during high-speed impact is assumed to be inviscid.

\begin{figure}
\centering
\includegraphics[width=9 cm]{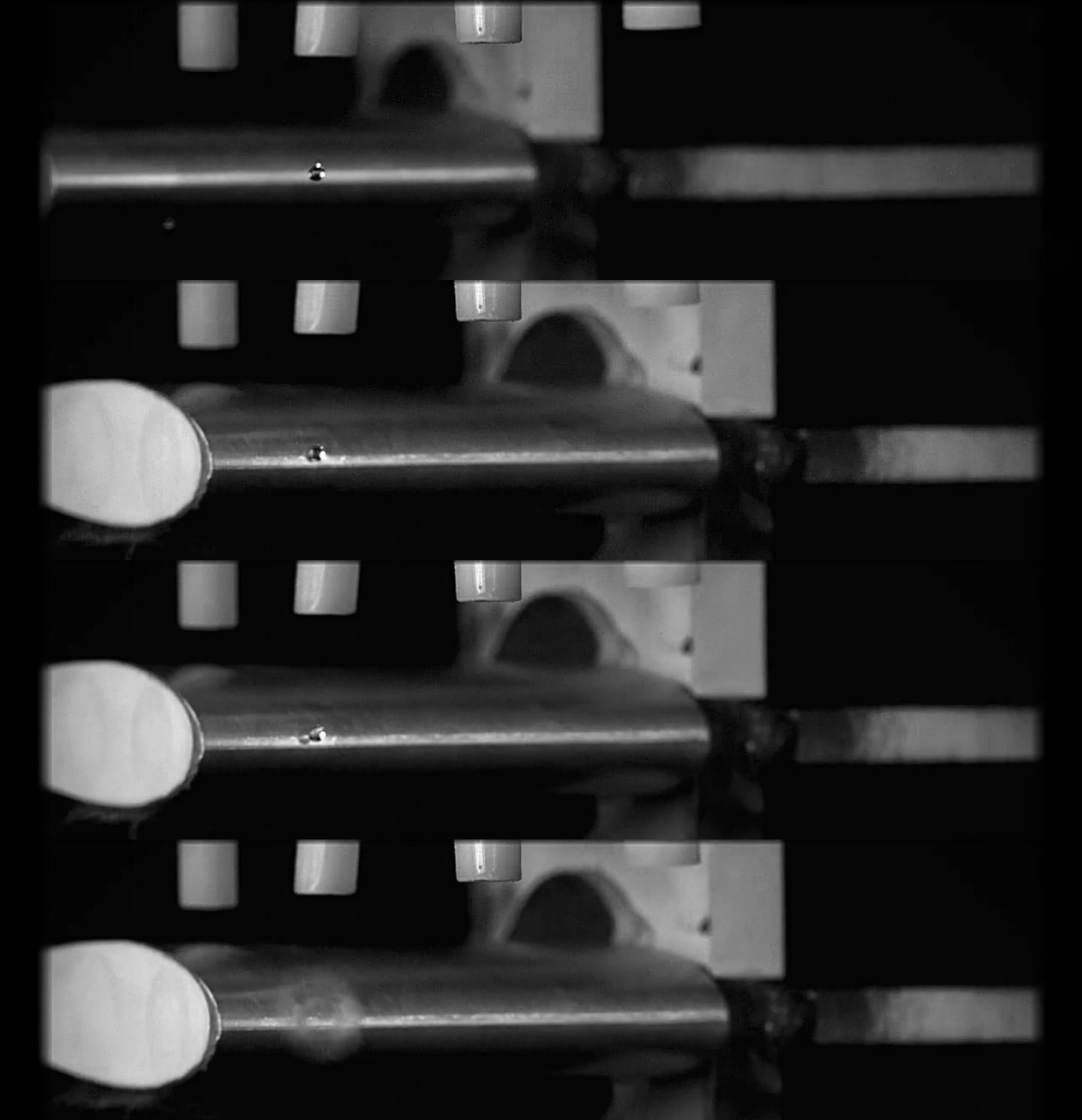}	
\caption{Whirling-arm rain erosion test (RET) \cite{RET}. From top to bottom: droplet free fall, droplet in front of the blade, droplet during impact, and droplet splashing after impact.}\label{figure_RET}
\end{figure}

In the context of rain erosion on wind turbine blades, where erosion results from repeated droplet impacts, the progression typically spans years. Consequently, fatigue models are often employed to estimate long-term damage accumulation and to guide blade maintenance. These include the widely used Springer model \cite{Springer1976,Fernando} and its subsequent modifications \cite{Eisenberg2018,Hao_field}, the kinetic energy approach \cite{Bech2018}, and the material strain-limit approach \cite{Verma2020}. Alternatively, industrial standards recommend estimating accumulated erosion damage using in-laboratory rain erosion tests (RET) \cite{RP-0171,RP-0573}. However, a significant mismatch has been reported between predictions from empirical fatigue models and RET results \cite{Herring}. A commonly cited reason is the idealised assumptions underlying empirical models, such as perfectly spherical droplets striking the surface perpendicularly \cite{Nick_Springer,Groucott}. In reality, due to aerodynamic effects around rotating blades, droplets are frequently distorted before impact and strike surfaces at arbitrary orientations and angles (see figure\ref{figure_RET}), both in laboratory experiments and in the field.

The present work contributes to rain erosion studies on wind turbine blades from a fluid-dynamic perspective, addressing the gap in the literature on distorted droplet impacts. We investigate liquid droplet impacts of arbitrary shapes and impact orientations prior to impact, and propose a \textit{universal scaling law} that enables predictions of impact loadings based solely on the initial droplet conditions. The focus is on the high-speed impact regime, where viscous and capillary effects\textemdash though potentially shape- and angle-dependent\textemdash are negligible. All flow parameters are carefully chosen to match with previous experimental studies on wind turbine blade erosion.

The paper is organised as follows. Section \ref{methods} introduces the mathematical models, with detailed descriptions of the computational framework for the fluid in section \ref{2.1} and for the solid in section \ref{2.2}. Section \ref{results} presents validation of the developed FSI model and the main findings, which are further discussed in section \ref{discussion}, leading to the proposed universal scaling law. The final section provides concluding remarks.

\section{Methods}
\label{methods}
A one-way adaptively coupled numerical fluid-structure interaction (FSI) model for high-speed multi-phase droplet impact evaluation is developed in the present work. An adaptive three-dimensional volume-of-fluid (VOF) simulation for incompressible multi-phase flow including liquid water and air components is used and described in Section \ref{2.1}. Following that, numerical pressure from the VOF fluid model is adaptively loaded as subroutines in finite element (FE) analysis on a wind turbine blade coating in Section \ref{2.2}.

\subsection{Multi-phase droplet impact model}
\label{2.1}
\subsubsection{Governing equations and dimensional scales}
\label{2.1.1}
\begin{figure}
\centering
\includegraphics[width=9 cm]{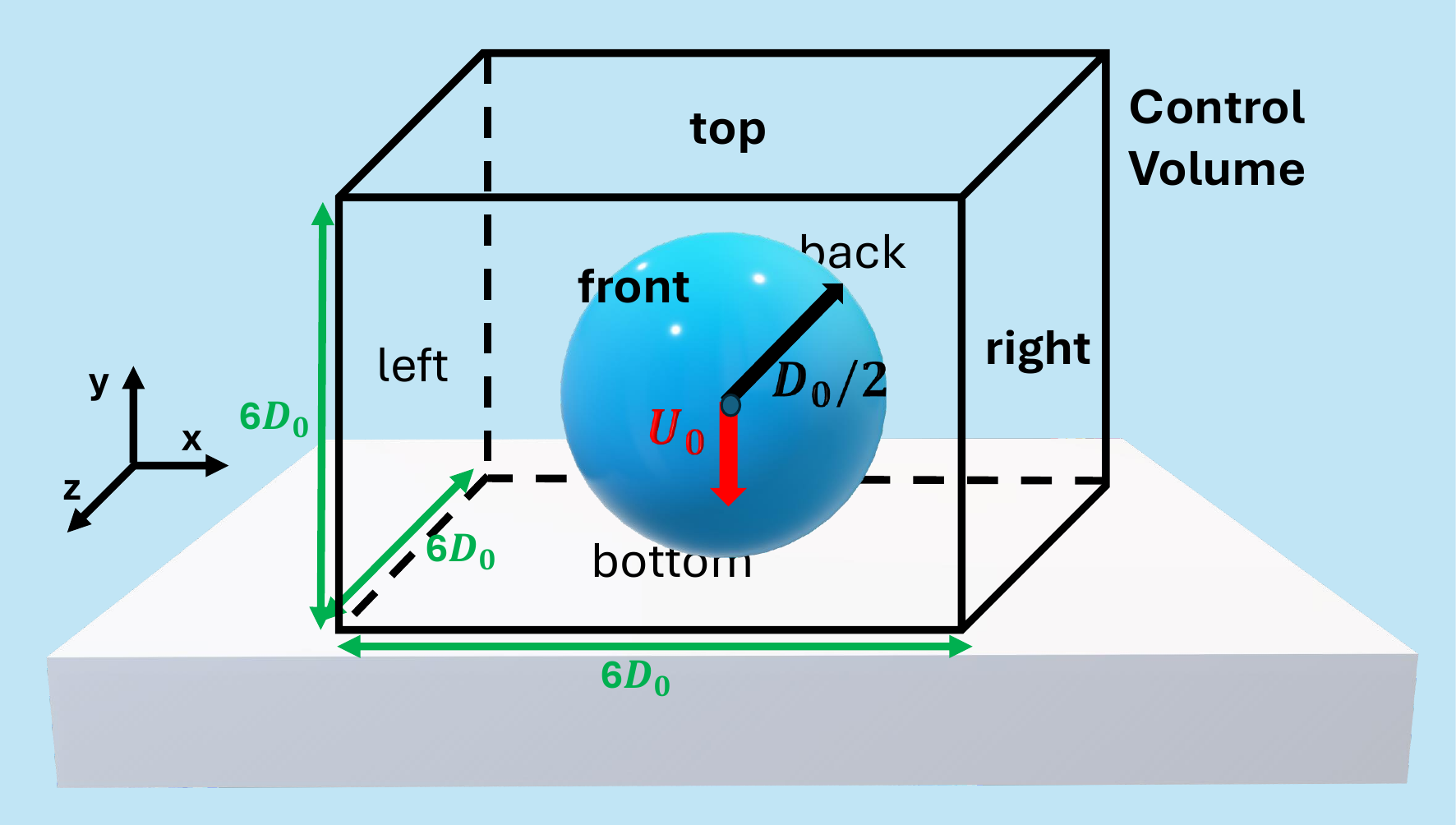}	
\caption{The domain of the three-dimensional VOF simulation including water droplet and air. The figure is for illustration purpose with lengths not to scale.}\label{figure_framework}
\end{figure}

In the context of raindrop impact on coated wind turbine blade surfaces, the considered geometry is shown in figure \ref{figure_framework} which consists of a liquid droplet impacting downwards onto an infinite, fixed and rigid solid surface in static ambient air. In the three-dimensional framework, we take the instant of impact as zero impact time $t=0$ and define the origin of the coordinate system at the center of the plate for all time $t \ge 0$. For the interest of the current study, it is reasonable to assume the model fluids, namely liquid droplets (taken to be water) and air, to be incompressible, viscous and immiscible for the entire impact process. Thereby, the model fluids satisfy the three-dimension incompressible momentum and continuity equations:
\begin{align}
  \frac{\partial \textbf{u}}{\partial t} + \nabla \cdot (\textbf{u} \times \textbf{u}) &= \frac{1}{\rho_i} [ -\nabla p + \nabla \cdot (2 \mu_i \textbf{D}) ] + \textbf{a}      \label{momentum} \\
\text{and}  \;\;\;\; \nabla \cdot \textbf{u}&=0 \label{continuity}
\end{align}
with the deformation tensor $\textbf{D}=[\nabla \textbf{u} + (\nabla \textbf{u})^T ]/2$ , where $\textbf{u}$ is flow speed, $p$ is pressure, $\mu$ is flow viscosity, $\rho$ is flow density,  $\textbf{a}$ is acceleration, and $\nabla$ is the nabla gradient operator in cartesian coordinate. Three-dimensional vectors are denoted by bold letters and subscripts $i$ denote water ($i=l$) or air ($i=g$). 

The fluid fields governing equations (\ref{momentum}) and (\ref{continuity}) are subject to the no-slip boundary condition at the rigid solid surface (denoted `bottom' in figure \ref{figure_framework}):
\begin{align}
   \textbf{u} \cdot \hat{\textbf{n}} = 0 \label{BC1}\\
\text{and}  \;\;\;\; \frac{dp}{dn}=0 \label{BC2}
\end{align}
where $\hat{\textbf{n}}$ is the unit normal vector at the boundary and $d/dn$ is the gradient operator in that normal direction. In a finite control volume (figure \ref{figure_framework}), the governing equations (\ref{momentum}) and (\ref{continuity}) also satisfy the free outflow boundary conditions:
\begin{align}
   \frac{d}{dn} \left( \textbf{u} \cdot \hat{\textbf{n}} \right) = 0  \label{BC3}\\
\text{and}  \;\;\;\; p=0 \label{BC4}
\end{align}
at the other five boundaries, namely the front, back, left, right and top surfaces of the control volume in figure \ref{figure_framework}.

Throughout this paper, variables are made dimensionless using $D_0$, $U_0$ and $\Pi_l$, with $D_0$ the droplet initial diameter, $U_0$ the impact speed and $\Pi_l$ the liquid density; and lower-case letters denote dimensionless forms of corresponding upper-case variables (except density) unless stated otherwise. Parameters of the model fluids, and corresponding dimensionless values, in the context of raindrop impacting on wind turbine blade surfaces are summarised in Table \ref{table_fluids} for water droplet and air at room temperature. The impact speed has been selected to be typical raindrop impact speeds in the research community on wind turbine blade leading-edge erosion (LEE)~\cite{Nick2023, Nils2023}, and the droplet diameter is selected for validation purpose to coincide with the experimental investigations~\cite{XJ} on water droplet impact. Based on the characteristic values of water in Table \ref{table_fluids}, we have the dimensionless parameters of the considered conditions: Reynolds number $Re=2.7\times10^5$, Weber number $We=3.7\times10^5$ and Froude number $Fr=614$.

\begin{table}
\caption{Properties and impact parameters for the fluid phases. G is gravitational acceleration and $\Sigma$ is surface tension between water and air.}\label{table_fluids}
\centering
\begin{tabular}{ccc} 
\toprule
\textbf{Variable letter} &  \textbf{Dimensional value}&  \textbf{Dimensionless value}\\
\midrule
$U_0$	&  100 (\unit{\meter\per\second}) & 1\\
\midrule
$D_0$	&  2.7 (mm) & 1\\
\midrule
$\Pi_l$ &  996 (\unit{\kilogram\per\cubic\meter}) & 1\\
\midrule
$\Pi_g$ &  1.29 (\unit{\kilogram\per\cubic\meter}) & 1.30e-3\\
\midrule
$M_l$ & 1e-3 (\unit{\kilogram\per\meter\per\second}) & 3.72e-6\\
\midrule
$M_g$ &  18.1e-6 (\unit{\kilogram\per\meter\per\second}) & 6.73e-8\\
\midrule
$G$ &  9.81 (\unit{\meter\per\second\squared}) & 2.65e-6\\
\midrule
$\Sigma$ &  73e-3 (\unit{\newton\per\meter}) & 2.71e-6\\
\bottomrule
\end{tabular}
\end{table}

\subsubsection{Volume-of-fluid fluids simulation}
\label{2.1.2}
Numerical computations for the three-dimensional multi-phase impact problem in Section \ref{2.1.1} are carried out in the open-source software \textit{Basilisk.fr}~\cite{Popinet2003, Popinet2009}. Basilisk has accurate solvers for incompressible Navier-Stokes equations (and other multi-physics solvers) using the finite volume (FV) method and a VOF approach to represent the multi-phase interfaces. The software is ideal for the current multi-phase high-speed impact problem framework since it adopts an adaptive octree spatial discretisation (in three dimension) which efficiently saves computational efforts while maintain high resolution at the fluid-fluid interfaces. Besides, the software is especially robust for surface-tension-driven interfacial flows due to its integrated VOF interface representation, balanced-force continuum-surface-force surface-tension formulation and height-function curvature estimation. This is helpful in applying the numerical model from low-Weber-number framework, such as our validation case~\cite{XJ}, to high-Weber-number framework in Section \ref{2.1.1}. Throughout the simulations, the schemes have a second-order convergence  in space and time on regular and dynamically refined grids.

For the two-phase flow in the current study, VOF adopts liquid fraction $f$ representing the liquid (water in this case) volume percentage in each cell. In other words, $f=0$ indicates the cell is occupied fully by gas while $f=1$ indicates the cell is occupied fully by water, and for any cell with $0 \leq f \leq 1$, there indicates an interface between gas and water. Readers are referred to Popinet~\cite{Popinet2009} and others~\cite{Rudman2005, Cummins2005} for details on re-constructuring the gas-water interface using liquid fraction and height functions. With liquid fraction $f$, Equation \ref{momentum} is re-written as:
\begin{align}
  &\frac{\partial \textbf{u}}{\partial t} + \nabla \cdot (\textbf{u} \times \textbf{u})  \nonumber \\ 
&= \frac{1}{\rho_l(1-f)+\rho_g f}[ -\nabla p + \nabla \cdot (2 (\mu_l (1-f)+\mu_g f) \textbf{D}) ] + \textbf{a}      \label{momentum2} 
\end{align}
at each cell to cater both phases. Additionally, the evolution of the liquid fraction $f$ is governed by the transport equation in time and space:
\begin{equation}
  \frac{\partial f}{\partial t} + \textbf{u} \cdot \nabla f = 0 \label{transport}
\end{equation}
In this way, Basilisk.fr solves Equations \ref{continuity}, \ref{momentum2} and \ref{transport}  subject to boundary conditions \ref{BC1}, \ref{BC2}, \ref{BC3} and \ref{BC4} in a simulation domain represented as the control volume in figure \ref{figure_framework}, which is a cube with dimension $6D_0$. The simulation uses a dimensionless unit system with fluids parameters listed in the third-column of Table \ref{table_fluids}.

In particular, we also elaborate the significance of the acceleration term $\textbf{a}$. In the case of the current study, it is consisted of the body force of gravity acceleration $\textbf{g}$ and interfacial acceleration $\textbf{a}_\textbf{f}$ at the interface. The later is defined in Basilisk as:
\begin{equation}
  \textbf{a}_\textbf{f} = \sigma \, \textbf{n} \, \delta_s / \rho
\end{equation}
where $\delta_s$ is the interface Dirac function that guarantees the interfacial acceleration term is only applied at cells containing interfaces, and $\rho$ is the density of the interface cell. Using a Continuum Surface Force (CSF) approximation, this can be expressed through the liquid fraction $f$:
\begin{equation}
  \textbf{a}_\textbf{f} = \sigma \, \nabla f /  (\rho_l(1-f)+\rho_g f)
\end{equation}
It is wroth-noting that, though the calculated dimensionless numbers $Re$, $Fr$ and $We$ are sufficiently large enough to neglect the presence of viscosity, gravity and surface tension for such high-speed impact, we still consider all of these factors in the current model. This is because the dimensionless numbers are indicators of the problem framework representing bulk water flow, while local regions or cells may have very distinctive dimensional scales. For instance, in the spreading sheet upon impact, Reynolds and Weber numbers may become locally small so that viscosity and surface tension can have a significant effect~\cite{Riboux2014}, for example, for secondary droplets from sheet atomisation upon splashing~\cite{Yarin2006, Papageorgiou}, Froude and Weber numbers may become locally small enough to consider gravity and surface tension. Besides, for the validation case in~\cite{XJ}, water droplet impacts at speed of $U_{\text{exp}}=2.67$ \unit{\meter\per\second} (while other physical parameters are kept the same as Table \ref{table_fluids}). This produces lower dimensional scales: $Re=7.2\times 10^3$, $We=263$ and $Fr=16$, all of which are close to the margins to be safely excluded~\cite{Gordillo2015}.

\begin{figure*}[t]
    \centering
    \begin{subfigure}[b]{0.32\textwidth}
        \centering
        \includegraphics[width=6.4 cm]{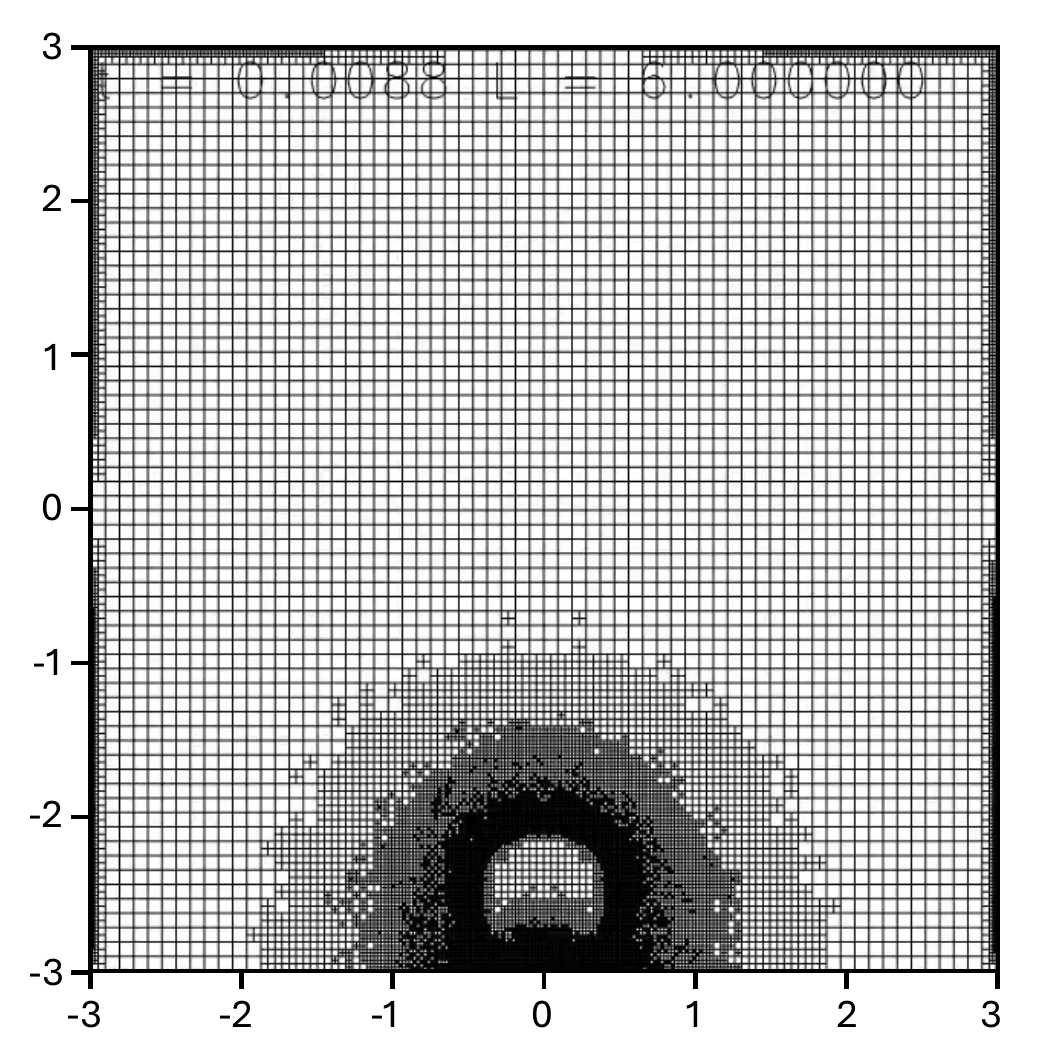}
        \caption{}
        \label{figure_AMR1}
    \end{subfigure}
    \begin{subfigure}[b]{0.34\textwidth}
        \centering
        \includegraphics[width=6.4 cm]{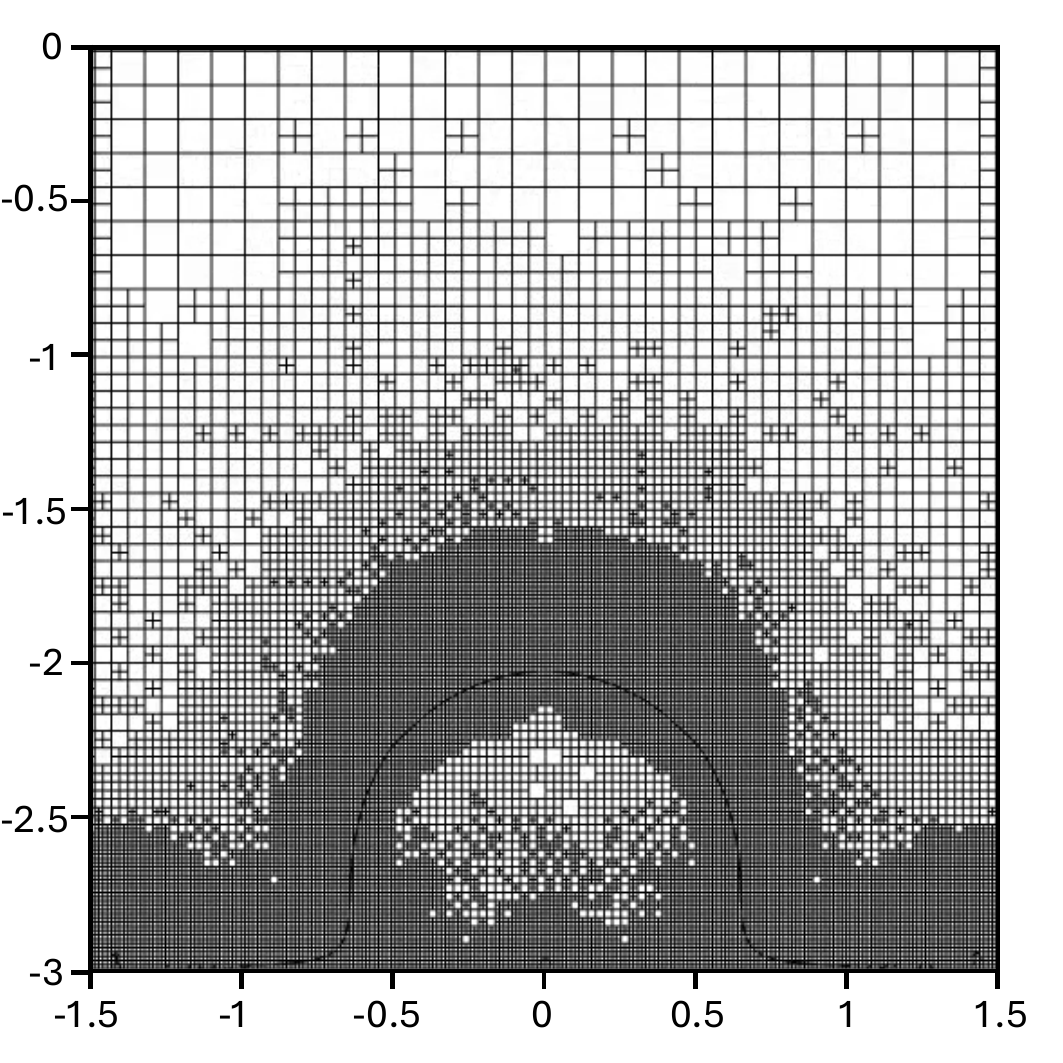}
        \caption{}
        \label{figure_AMR2}
    \end{subfigure}
    \begin{subfigure}[b]{0.32\textwidth}
        \centering
        \includegraphics[width=6.4 cm]{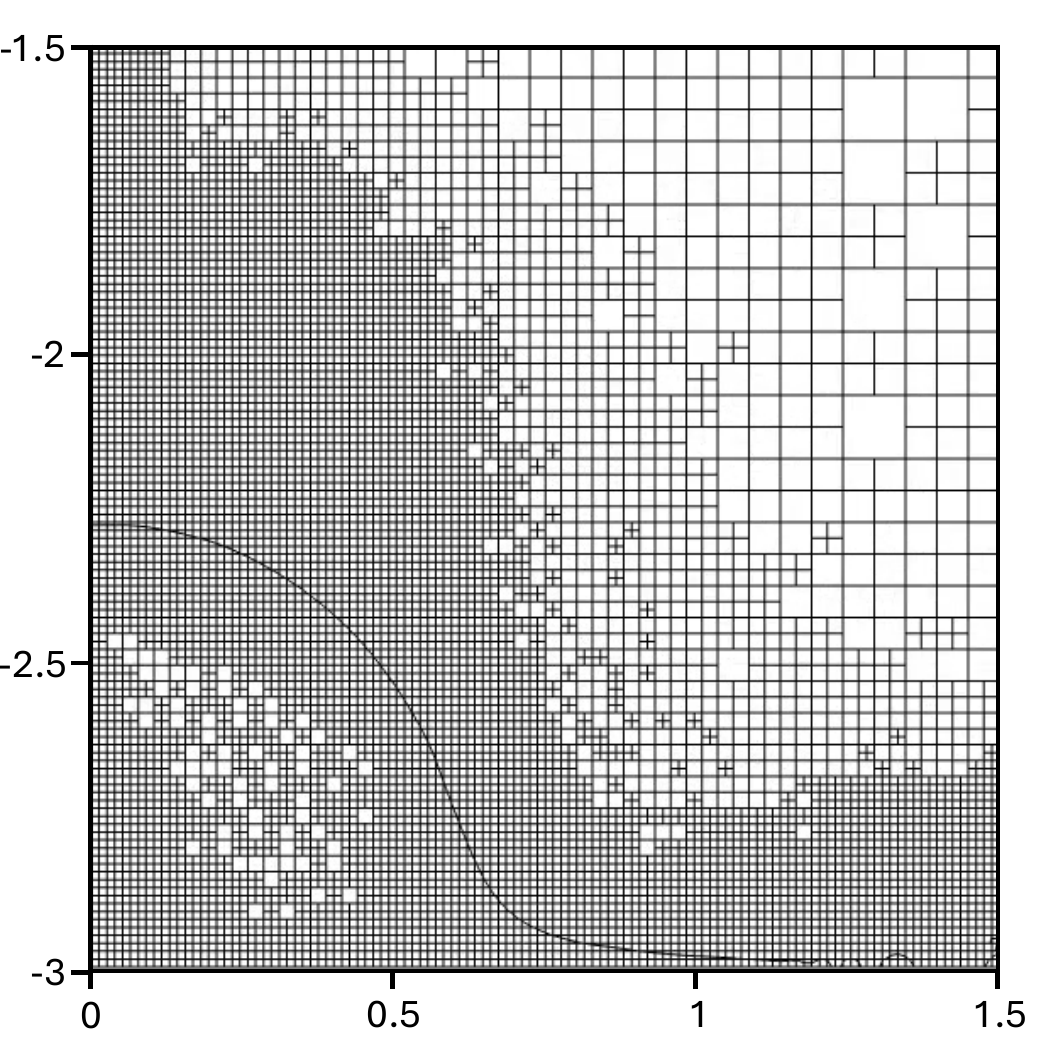}
        \caption{}
        \label{figure_AMR3}
    \end{subfigure}
    \caption{The adaptive mesh refinement (AMR) scheme for spatial discretisation of the VOF simulation, at mesh resolutions between \textit{Levels} $6-9$. The snapshots are taken at dimensionless times ($a$) $t=0.0088$, ($b$) $t=0.2156$, and ($c$) $t=0.3432$, respectively, as representatives of early, intermediate, and late impact times.}
\label{figure_AMR}
\end{figure*}

The three-dimensional numerical model is solved on a simulation domain of dimensionless size $6^3$ with total simulation time of $t_{end}=1.54$ at a timestep of $\Delta t=4.4\times10^{-3}$. The timestep is selected to satisfy the Courant-Friedrichs-Lewy (CFL) condition of computational fluid dynamics (CFD) simulations, and is also aligned with the timestep constraint of the numerical material analysis outlined in Section \ref{2.2}. The total simulation time, $t_{end}$, defines the length of the simulation domain based on the spreading law $2\sqrt{6t_{end}}\approx6$~\cite{Wagner,Riboux2014}. The spatial discretisation of the current model uses an adaptive mesh refinement (AMR) starting from \textit{Level} $9$ (to ensure smooth initial interface), and varies between \textit{Levels} $6-9$ throughout the simulation, where level $L$ is defined in Basilisk as cubic cells (in three dimensions) of dimensionless side length $6/2^L$. Figure \ref{figure_AMR} presents the adaptive mesh at three representative dimensionless times: early time $t=0.0088$ (figure \ref{figure_AMR1}, which is merely the second time step after the impact, showing the initial mesh degradation from the starting condition of Level $9$ in the entire simulation domain), intermediate time $t=0.2156$ (figure \ref{figure_AMR2}, which is zoomed-in twofold around the droplet, showing the finest mesh at fluid-fluid interfaces and a `thick' range of their neighbours), and late time $t=0.3432$ (figure \ref{figure_AMR3}, which is zoomed-in fourfold on the right half of the droplet), for the $x$-$y$ plane of the three-dimensional simulation at $z=0$. The finest mesh level is selected to align with the guidance in the community of CFD simulation on droplet impact that droplet diameter shall span $\mathcal{O}$(100) cells \cite{Tsinghua2018}. Thanks to the smartly selective AMR scheme, the solver automatically (based on specified tolerance conditions) downgrades, or upgrades, mesh levels at regions that contain delicate, or crude, features of the fluid flows, so users do not have to maintain the required  finest levels for all simulation time. The adaption of AMR effectively keeps the numerical model at a manageable size while not sacrificing results reliability. In the current study, the criteria for AMR is based on a tolerance of $10^{-4}$ on calculated $f$ and $\textbf{u}$ at each timestep, which is one order less than the recommended tolerance value by Popinet~\cite{Popinet2009} on the divergence of the velocity field. This mesh setting leads to the total number of cells for the entire simulation domain around $\mathcal{O}(10^{6-8})$. A typical simulation of the developed VOF model with the above numerical conditions takes about $10^2$ CPU hours to solve on a AMD EPYC 7742\raisebox{1ex}{\scriptsize TM} processor with 1000 GB of RAM.

\subsection{Finite element model of blade coating material}
\label{2.2}
\begin{figure}
\centering
\includegraphics[width=9.5 cm]{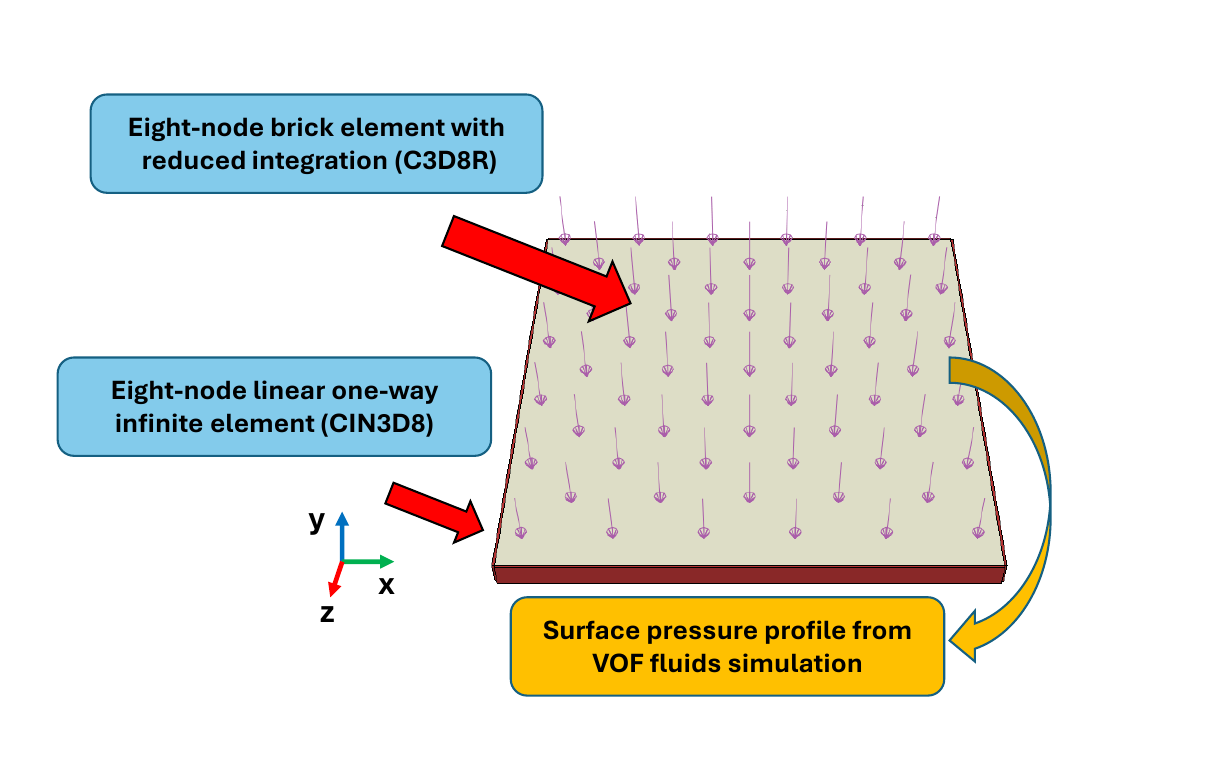}	
\caption{The geometry and material elements in the FE model. The labelled element codes are used in commercial software ABAQUS.}\label{figure_NNCM}
\end{figure}

Wind turbine blade coating is modelled in commercial software ABAQUS\raisebox{1ex}{\scriptsize \textregistered} using three-dimensional FE analysis. Blade materials are modelled by the eight-node brick element with reduced integration (C3D8R) with a layer of one-way infinite elements (CIN3D8) on four sides and bottom to eliminate wave reflections due to the finite geometry. The development of one-way FSI coupling framework adaptively combines numerical pressure from the bottom surface of the VOF simulations (figure \ref{figure_framework}) as a time- and spatial-dependent loading on the surface of the blade coating material to model the impact loading of a liquid droplet. It is noted that the VOF fluids model assumes rigid bottom surface for the entire simulation durations, while coating material in the FE analysis dynamically deforms and recovers subject to impact loadings. Nevertheless, it has been found that surface motions upon droplet impact has negligible effect in one-way coupled FSI system for materials with Young's modulus greater than the stiffness threshold around $50,000$ \unit{\pascal}~\cite{Hao_softness}. While general wind turbine blade protection coatings (including the coating material used in the current study) use materials similar to the matrix material, such as epoxy, polyester or polyurethane~\cite{Herring2019}, which typically have Young's modulus of $\mathcal{O}(10^{9-12})$ \unit{\pascal} well above the stiffness threshold, and are therefore safe to be modelled as rigid surfaces in the fluids simulation. The user-defined loading is achieved by the VDLOAD-subroutine in ABAQUS for explicit dynamic analysis. Figure \ref{figure_NNCM} displays the geometry and FE elements in the study. Four sides and bottom of the geometry are fixed by encaster boundary conditions to model the geometric constraints of material elements in real wind turbine blades. 

The three-dimensional FE model is solved on a substrate of size 16.2 mm (length) $\times$ 16.2 mm (width) $\times$ 4 mm (height) so that the surface size is in line with the VOF simulation (figure~\ref{figure_framework}). The total simulation time and the timestep are also kept the same as those in the VOF simulation. This timestep agrees with the guidance in the field of droplet impact erosion that peak pressure force shall be captured in no less than ten steps~\cite{Doagou}, and our sensitivity check indeed shows no results difference at finer timesteps. The substrate is bias-meshed with finer mesh towards the impact surface (mesh size $125$ \unit{\micro\meter}) and coarser at bottom (mesh size $200$ \unit{\micro\meter}). This mesh setting leads to the total number of elements (including the one-way infinite layers) in excess of $4\times10^{5}$, which has been proven efficient and successful in the validation of a previous study~\cite{Hao_ANCM} against experimental data. A typical simulation of the three-dimensional numerical FE model for material cases of the current study takes about $50$ CPU hours to solve on a AMD Ryzen Threadripper PRO 5975WX\raisebox{1ex}{\scriptsize TM} processor with 270 GB of RAM. 

In the current study, we use \textit{3M}\raisebox{1ex}{\scriptsize TM} Wind Blade Protection Coating $W4600$~\cite{3M} as a typical blade coating material in the FE analysis. Coating material properties are listed in Table \ref{table_mat}, and by theory of elasticity~\cite{Timoshenko}, Young's modulus of the coating is calculated as:
\begin{equation}
  E_c = \frac{\rho_c C_c^2 (1+\nu_c) (1-2\nu_c)}{1-\nu_c}=10.3 \,\text{GPa}
\end{equation}
Thereby, in the FE material analysis, it is reasonable to assume a linear elastic model for the simulated coating material since impact by single water droplet of conditions in Table \ref{table_fluids} can merely deform such high-stiffness material elastically.

\begin{table}
\caption{Properties for the commercial wind turbine blade coating material, 3M W4600~\cite{Julie2020}.}\label{table_mat}
\centering
\begin{tabular}{ccc} 
\toprule
\textbf{Sonic Speed $C_c$} &  \textbf{Density $\rho_c$}&  \textbf{Poisson's Ratio $\nu_c$}\\
\midrule
1900 (\unit{\meter\per\second})	& 1100 (\unit{\kilogram\per\cubic\meter})& 0.3\\
\bottomrule
\end{tabular}
\end{table}

\section{Results}\label{results}
This section first examines the grid independence of the numerical simulations and validates the developed fluid–structure interaction (FSI) model against experimental measurements. A parametric investigation is then conducted to elucidate the effects of droplet geometry on the impact loading of the solid surface. Specifically, the effects of droplet shape, rotation angle, and volume are investigated through analyses of the transient pressure distributions and impact forces. The results are further interpreted using theoretical considerations, particularly momentum conservation and scaling relationships, to explain the observed differences in impact responses.

\begin{figure}
\centering
\includegraphics[width=9.5 cm]{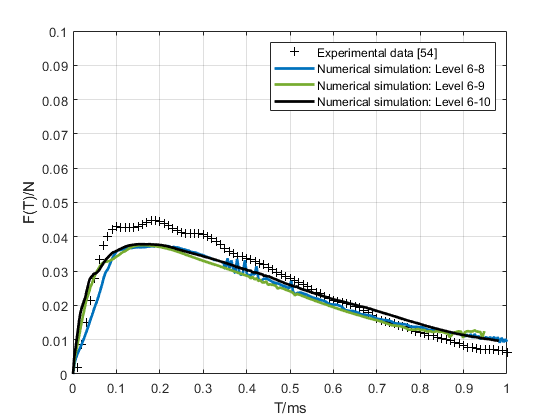}	
\caption{Impact force temporal evolution at different mesh resolutions. The experimental data from \cite{XJ} is also superimposed.}\label{force_ref}
\end{figure}

\subsection{Grid-independence test and model validation}\label{3.1}
To validate the developed one-way adaptively coupled FSI model, we have tailored the numerical parameters to simulate an experiment in the literature~\cite{XJ}. Therein, a schematic of experimental setup was consisted of a transducer attached below a machined and polished aluminium plate, together with amplifier and high speed camera to filter and acquire data respectively. Droplets, with diameters of around $2.7$ mm and impact speeds $2.67$ \unit{\meter\per\second}, were generated under gravity by high-precision syringe pump connected to a flat-tipped needle. We refer to Zhang \textit{et al.}~\cite{XJ} for further experimental details. The impact parameters for the validation case are listed in Table \ref{table_val}, and we note that, only in this validation Section \ref{3.1}, we made variables dimensionless using the characteristic speed $U_{\text{exp}}$ instead of $U_0$ from Table \ref{table_fluids}. Though $U_{\text{exp}}$ is two-order of magnitudes smaller than $U_0$, this validation case is one of a few experimental datasets available in the literature, which are all conducted at impact speeds of order $\textit{O}(1 \unit{\meter\per\second})$; besides, thanks to the inclusion of the capillary, viscous, and gravitational effects in Section \ref{2.1.2}, the VOF model is applicable to this relatively lower impact speed. Material properties of the aluminium plate used in experiments are listed in Table \ref{table_Al}. The aluminium material is stiffer than the 3M coating, thereby all previous assumptions from Section \ref{2.2} are still valid, namely a rigid surface in the fluids model and a linear elastic material model in FE analysis.

\begin{table}
\caption{Properties and impact parameters of the fluid phases used in the experiments~\cite{XJ}. Note that variables in this evaluation case are made dimensionless using $U_{\text{exp}}$ for characteristic velocity.} \label{table_val}
\centering
\begin{tabular}{ccc} 
\toprule
\textbf{Variable letter} &  \textbf{Dimensional value}&  \textbf{Dimensionless value}\\
\midrule
$U_{\text{exp}}$	&  2.67 (\unit{\meter\per\second}) & 1\\
\midrule
$D_0$	&  2.7 (mm) & 1\\
\midrule
$\Pi_l$ &  996 (\unit{\kilogram\per\cubic\meter}) & 1\\
\midrule
$\Pi_g$ &  1.29 (\unit{\kilogram\per\cubic\meter}) & 1.30e-3\\
\midrule
$M_l$ & 1e-3 (\unit{\kilogram\per\meter\per\second}) & 1.39e-4\\
\midrule
$M_g$ &  18.1e-6 (\unit{\kilogram\per\meter\per\second}) & 2.52e-6\\
\midrule
$G$ &  9.81 (\unit{\meter\per\second\squared}) & 3.71e-3\\
\midrule
$\Sigma$ &  73e-3 (\unit{\newton\per\meter}) & 3.80e-3\\
\bottomrule
\end{tabular}
\end{table}

\begin{table}[t]
\caption{Properties for the aluminium material used in the experiments~\cite{XJ}.}\label{table_Al}
\centering
\begin{tabular}{ccc} 
\toprule
\textbf{Young's Modulus $E_{Al}$} &  \textbf{Density $\rho_{Al}$}&  \textbf{Poisson's Ratio $\nu_{Al}$}\\
\midrule
70 (GPa)	& 2820 (\unit{\kilogram\per\cubic\meter})& 0.3\\
\bottomrule
\end{tabular}
\end{table}

For the incompressible VOF simulation, the grid independence is analysed by choosing three different upper grid resolutions in the adaptive mesh refinement scheme: Level $8$, Level $9$, and Level $10$ with the same lower grid resolution of Level $6$. These correspond to the finest mesh sizes of $2.34\%D_0$, $1.17\%D_0$, and $0.59\%D_0$, respectively. All adhere to the mesh size guidance in the literature for Computational Fluid Dynamics (CFD) simulation on droplet impact, that droplet diameter spans $\mathcal{O}(100)$ cells~\cite{Tsinghua2018}. Under the three mesh resolutions, simulation results of the impact forces on the solid surface are shown in figure \ref{force_ref}. In all three cases, the temporal evolution trends of the impact force are well-reproduced: a rapid growth at the early stage of the impact to the force maximum, after which the force gradually decreases to zero. Particularly, we observe that all three cases consistently reach the force maximum at time around $0.15$ ms. Nevertheless, at the mesh resolution of Levels $6-8$ (blue line), the impact force growth at the early stage of the impact, which plays a key role in impact erosion, is notably slower than the other cases and deviates from the experimental data (described below) in figure \ref{force_ref}. This is due to the relatively coarser initial spherical geometry approximated by the lower-resolution meshes before impact \cite{Hao_ana}, while the rapid force growths at mesh resolutions of Levels $6-9$ (green line) and Levels $6-10$ (black line) collapse onto a consistent curve. Hereby, to balance the computational efficiency and the delicate accuracy of material elastic waves, the mesh resolution of Levels $6-9$ (mesh size of $1.17\%D_0$) is chosen in the present study.

To validate the numerical FSI model, we superimpose the experimental data from Zhang \textit{et al.}~\cite{XJ} in figure \ref{force_ref}. As shown, the simulated impact force (at Levels $6-9$) generally agrees well with the experimental measurements, except for a slight discrepancy near the force maximum. The experimental data reaches its maximum value, $F_{\text{max}}=0.045\,N$, at $T=0.185$ ms, whereas the numerical simulation peaks at $0.038\,N$. Notably, Zhang \textit{et al.}~\cite{XJ} also reported a similar underestimation in their finite-element simulations, albeit with a smaller deviation. Considering the overall temporal evolution of the impact force, particularly the rapid load increase in the early impact stage, the present model yields predictions that are consistent with the experimental observations within an acceptable margin of accuracy.

\subsection{Parametric investigation}\label{3.2}

\begin{figure}
\centering
\includegraphics[width=9.5 cm]{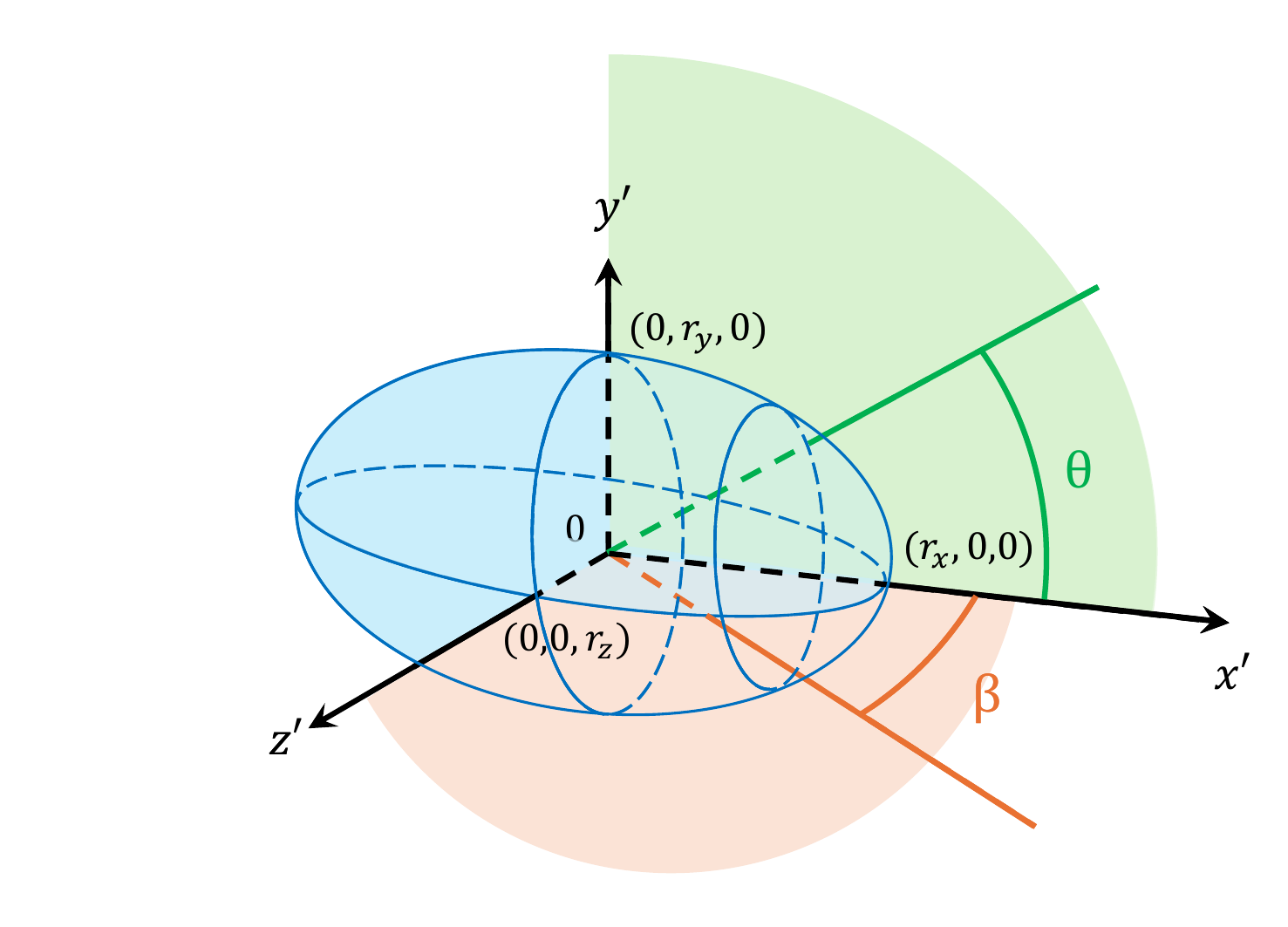}	
\caption{The general ellipsoid in the three-dimensional cartesian coordinate with its center coincided with the coordinate origin. The ellipsoid has radii, $r_x$, $r_y$, and $r_z$ in $x-$, $y-$, and $z-$directions, respectively.}\label{parametric}
\end{figure}

To investigate the effects of droplet shape and rotation angle on material damages, we investigate the ellipsoid of general three-dimensional shape in figure \ref{parametric}. The ellipsoid has radii $r_x$, $r_y$, and $r_z$ in $x'$-, $y'$-, and $z'$-direction, respectively, with its centre aligned with the coordinate centre. Here, the prime superscript indicates coordinates pre-rotation. This gives the shape of the ellipsoid as:
\begin{equation}
\left(\frac{x'}{r_x}\right)^2+\left(\frac{y'}{r_y}\right)^2+\left(\frac{z'}{r_z}\right)^2=1
\end{equation}
Based on this reference case, rotation angles in the $x$-$y$ plane is denoted as $\theta$ (anti-clockwise from the positive $x$-axis), and in the $z$-$x$ plane is denoted as $\beta$ (clockwise from the positive $x$-axis), as illustrated in figure \ref{parametric}. This gives the axes rotation (\textit{w.l.o.g.} $\beta=0$):
\begin{equation}
\begin{pmatrix}
x'\\
y'-r_y\\
z'
\end{pmatrix}=
\begin{pmatrix}
cos(\theta) & sin(\theta) & 0\\
-sin(\theta) & cos(\theta) & 0\\
0 & 0 & 1
\end{pmatrix}
\begin{pmatrix}
x\\
y\\
z
\end{pmatrix}
\end{equation}
Hence, the shape of the ellipsoid, in $x-y-z$ coordinate, is described as:
\begin{equation}
\left(\frac{xcos(\theta)+ysin(\theta)}{r_x}\right)^2+\left(\frac{-xsin(\theta)+ycos(\theta)}{r_y}\right)^2+\left(\frac{z}{r_z}\right)^2=1
\label{eq1}
\end{equation}
Finally, we offset the ellipsoid along the positive $y$-axis by distance $r_{\text{max}}$ so that the droplet is above the surface at initial simulation condition (see figure \ref{figure_framework}), where $r_{\text{max}}$ is the bigger value among $r_x$  and $r_y$. In this way, $y$ should be replaced by $y-r_{\text{max}}$ in equation \ref{eq1}, as:
\begin{align}
\left(\frac{xcos(\theta)+(y-r_{\text{max}})sin(\theta)}{r_x}\right)^2+&\left(\frac{-xsin(\theta)+(y-r_{\text{max}})cos(\theta)}{r_y}\right)^2\nonumber \\ 
&\;\;\;\;\;\;\;\;\;\;\;\;\;\;\;+\left(\frac{z}{r_z}\right)^2=1
\label{ellipsoid}
\end{align}
We note that, by doing so, the droplet after rotation does not guarantee a `touch' initial condition between the droplet and the surface, and there may exist a gap in the middle.

\begin{table*}
\caption{Summary of the investigated cases in the parametric study, with their radii $r_x$, $r_y$, and $r_z$, rotation angles $\theta$ and $\beta$, and liquid volume.}\label{table_para}
\centering
\begin{tabular}{cccccccc} 
\toprule
 & \textbf{Case $1$} & \textbf{Case $2$} & \textbf{Case $3$} &  \textbf{Case $4$} &  \textbf{Case $5$} &  \textbf{Case $6$} &  \textbf{Case $7$}\\
& \textbf{Sphere} & \textbf{Ellipsoid} & \textbf{Spheroid-0} &  \textbf{Spheroid-45} &  \textbf{Spheroid-75} &  \textbf{Spheroid-90} &  \textbf{Sphere-s}\\
\midrule
\textbf{$r_x$} &  $1/\sqrt[3]{2}$ &$1/\sqrt{2}$&$1$&$1$&$1$&$1$&$1/2$\\
\midrule
\textbf{$r_y$} &  $1/\sqrt[3]{2}$ &$1/\sqrt{2}$&$1/2$&$1/2$&$1/2$&$1/2$&$1/2$\\
\midrule
\textbf{$r_z$} &  $1/\sqrt[3]{2}$ &$1$&$1$&$1$&$1$&$1$&$1/2$\\
\midrule
\textbf{$\theta$} &$0^\circ$   &$0^\circ$  &$0^\circ$ &$45^\circ$ & $75^\circ$& $90^\circ$ &$0^\circ$\\
\midrule
\textbf{$\beta$} &$0^\circ$   &$0^\circ$  &$0^\circ$ &$0^\circ$ & $0^\circ$& $0^\circ$ &$0^\circ$\\
\midrule
\textbf{Total Volume} &  $3\pi/8$ &$3\pi/8$&$3\pi/8$&$3\pi/8$&$3\pi/8$&$3\pi/8$&$3\pi/32$\\
\bottomrule
\end{tabular}
\end{table*}

\begin{figure*}
    \centering
    \begin{subfigure}[b]{0.49\textwidth}
        \centering
        \includegraphics[width=9.5cm]{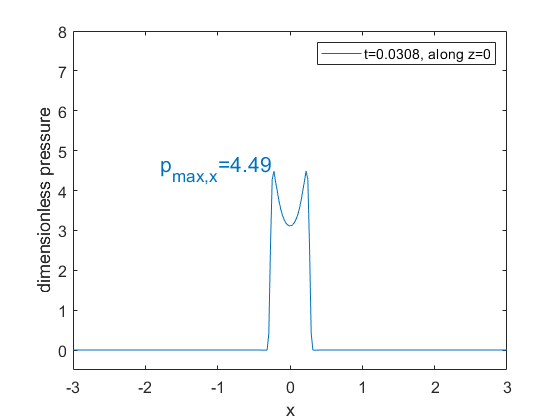}
        \caption{}
        \label{figure_PS1}
    \end{subfigure}
    \begin{subfigure}[b]{0.49\textwidth}
        \centering
        \includegraphics[width=9.5cm]{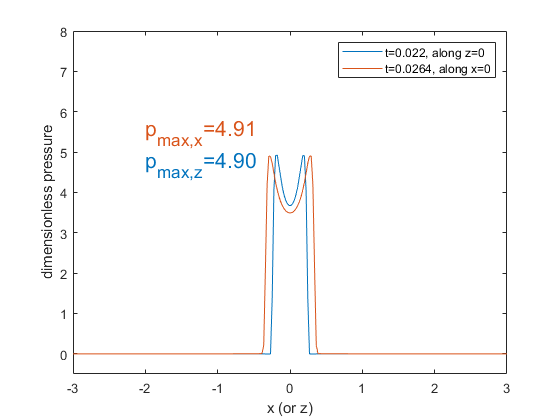}
        \caption{}
        \label{figure_PS2}
    \end{subfigure}
    \begin{subfigure}[b]{0.49\textwidth}
        \centering
        \includegraphics[width=9.5cm]{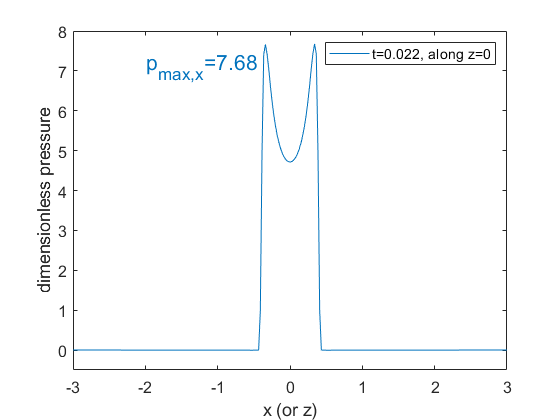}
        \caption{}
        \label{figure_PS3}
    \end{subfigure}
    \begin{subfigure}[b]{0.49\textwidth}
        \centering
        \includegraphics[width=9.5cm]{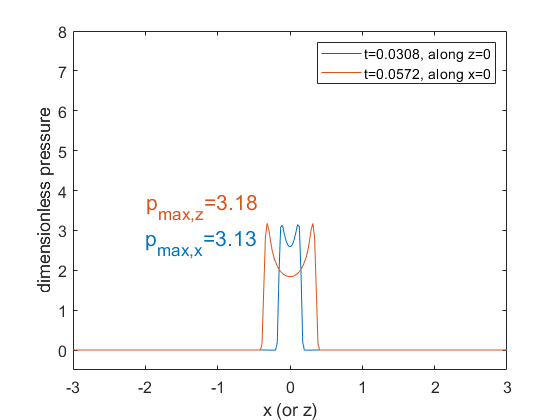}
        \caption{}
        \label{figure_PS4}
    \end{subfigure}
    \caption{Impact pressure distributions along the line $z=0$ (and the line $x=0$ for non-axisymmetric cases) on the solid surface in the four parametric cases (a) Sphere, (b) Ellipsoid, (c) Spheroid-0, and (d) Spheroid-90. Each distribution happens at the time of the pressure peak along the corresponding line.}
\label{figure_PS}
\end{figure*}

To isolate the effects of droplet shape and rotation angle, we pose a constraint on the radii $r_x$, $r_y$, and $r_z$ so that the total liquid volume, and hence the initial impact momentum, remains unchanged. This produces the different cases of table \ref{table_para}. Besides, we add the \textit{case 7: Sphere-s} to explore the effect of liquid volume while keeping its shape the same as \textit{case 1: Sphere}. From now on, we employ the impact conditions, and liquid and material properties as in tables \ref{table_fluids} and \ref{table_mat}.

\subsubsection{Effect of droplet shape}\label{3.2.1}
In this section, we compare the cases $1$, $2$, $3$ and $6$ of table \ref{table_para}. We note that: $1$) Spheroid-90 is equivalent to an ellipsoid of $r_x=1/2$, $r_x=1$, and $r_z=1$ with $\theta=0$; while $2$) Ellipsoid and Spheroid-0 are \textit{not} $\beta$- rotation-equivalent as their radii are not the same. Hereby, the cases of Sphere and Spheroid-0 are axisymmetric on the $x$-$z$ plane with respect to the $y$-axis, while the cases of Ellipsoid and Spheroid-90 are not.

\begin{figure}
\centering
\includegraphics[width=9.5 cm]{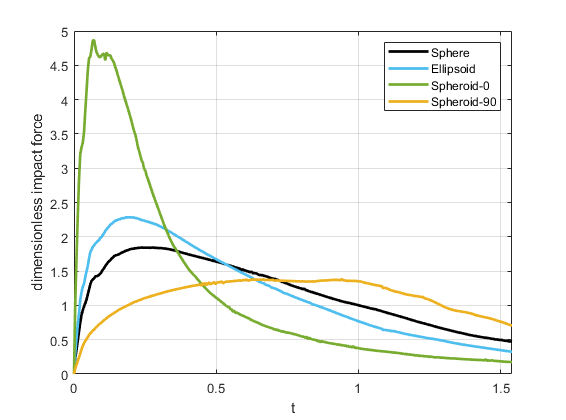}	
\caption{Impact force temporal evolution for the four parametric cases: Sphere, Ellipsoid, Spheroid-0, and Spheroid-90.}\label{force_PS}
\end{figure}

We present the maximum pressure distributions on the impact surface along the lines $x=0$ and $z=0$ (for non-axisymmetric cases) in figure \ref{figure_PS}. As shown, the pressure peaks earliest in the case of Spheroid-0 (figure \ref{figure_PS3}), which also exhibits the largest dimensionless pressure maximum, $p_{\text{max},x}=7.68$. This is followed by Ellipsoid (figure \ref{figure_PS2}) (with its pressure maximum happening along $x=0$) and then Sphere, both of which have comparable peak values. In contrast, Spheroid-90 (figure \ref{figure_PS4}) shows a noticeably smaller pressure maximum, occurring at a later time. 

We observe a similar trend for the impact force on the surface in figure \ref{force_PS}. Since force is the surface integral of pressure, this consistency indicates that the observed variation is a general characteristic of the impact pressure distribution over the surface, rather than being limited to the pressure maximum alone. Nevertheless, as shown in figure \ref{force_PS}, for cases that reach higher peak forces earlier, the subsequent decay stage after the peak force is also faster.

We integrate the force distributions over time, which equals the total momentum change during impact. The initial dimensionless (vertical) momentum, $m_{ini}$, of the four cases is:
\begin{equation}
    m_{ini}=v_0\;u_0=2\pi/3\approx2.09
\label{mom}
\end{equation}
where $v_0$ denotes the dimensionless liquid volume (of cases $1-6$). During impact events, all initial momentum eventually converts into radial momentum \cite{Egger2010}, and hence we expect the total momentum changes in all four cases to equal $2.09$ in theory. Indeed, the integrals provide $2.09$ for Sphere, $2.09$ for Ellipsoid, $1.98$ for Spheroid-0, and $2.10$ for Spheroid-90. This explains the trend observed in figure \ref{force_PS} that, for the cases which reach higher peak forces earlier, the subsequent decay stage after the peak force is faster so that the total momentum change, \textit{i.e.}, area under the force-time curve, is conserved.


\subsubsection{Effect of rotation angle}\label{3.2.2}
In this section, we compare the cases $3$, $4$, $5$ and $6$ of table \ref{table_para}, all of which have the same shape but different rotation angles, $\theta$. Rotation angle is different to impact angle, and the latter is always vertically downwards towards the surface with unit speed in the dimensionless system. Among the investigated cases of this section, only Spheroid-0 is axisymmetric on the $x$-$z$ plane with respect to the $y$-axis.

\begin{figure}
    \centering
    \begin{subfigure}[b]{0.49\textwidth}
        \centering
        \includegraphics[width=9.5cm]{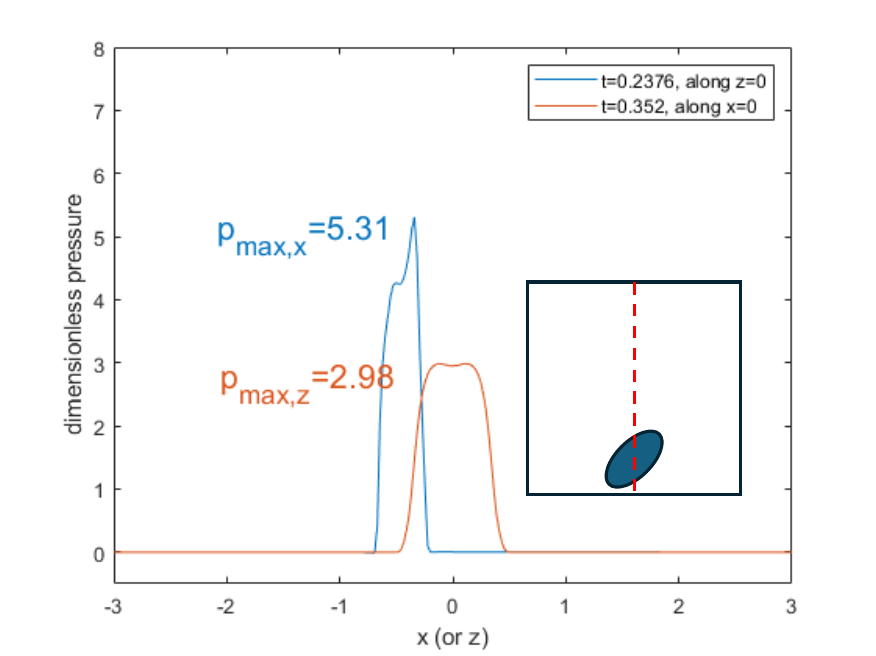}
        \caption{}
        \label{figure_PA1}
    \end{subfigure}
    \begin{subfigure}[b]{0.49\textwidth}
        \centering
        \includegraphics[width=9.5cm]{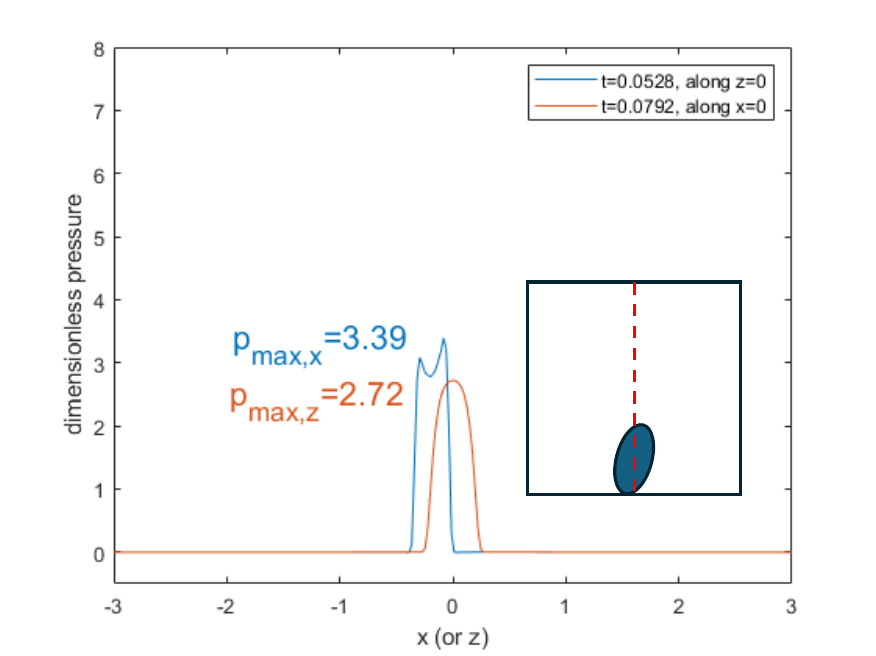}
        \caption{}
        \label{figure_PA2}
    \end{subfigure}
    \caption{Impact pressure distributions along the lines $z=0$ and $x=0$ on the solid surface in the non-axisymmetric cases (a) Spheroid-45 and (b) Spheroid-75. Each distribution happens at the time of the pressure peak along the corresponding line. Insets show the front view of the initial geometry at $x$-$y$ plane.}
\label{figure_PA}
\end{figure}

Figure \ref{figure_PA} presents the same information as figure \ref{figure_PS} but for the complementary cases of Spheroid-45 and Spheroid-75. As can be seen, for these two cases (only), the instants of pressure peaks along the lines $z=0$ and $x=0$ are notably distinctive. This unique feature is caused by the rotations, $\theta$, which move the lowest point of the initial geometry off the central $y$-axis (see the insets of figure \ref{figure_PA}). In other words, there experiences a delay of impact in the $y$-$z$ plane (red dashed lines). To facilitate a consistent comparison in the following analysis, this impact delay is accounted for and removed from the time histories.

Regarding the peak impact pressure distributions among the four cases, there is a monotonic trend of decreasing peak forces as the rotation angle, $\theta$,  increases, \textit{i.e.}, in the order of decreasing magnitudes: Spheroid-0 (figure \ref{figure_PS3}), Spheroid-45 (figure \ref{figure_PA1}), Spheroid-75 (figure \ref{figure_PA2}), and Spheroid-90 (figure \ref{figure_PS4}). Besides, the two middle cases of Spheroid-45 and Spheroid-75 exhibit asymmetric pressure distributions along the $x$-axis since the droplet geometries are asymmetric (in the $x$-$z$ plane) pre-impact.

\begin{figure}
\centering
\includegraphics[width=9.5 cm]{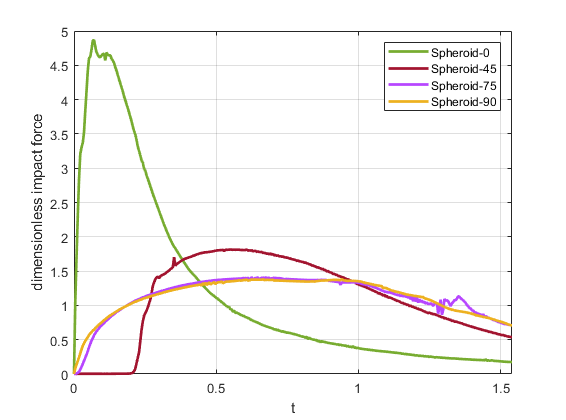}	
\caption{Impact force temporal evolution for the four parametric cases: Spheroid-0, Spheroid-45, Spheroid-75, and Spheroid-90.}\label{force_PA}
\end{figure}

Impact forces on the surface are shown in figure \ref{force_PA}. Because of the aforementioned gap between the droplet and surface (as the way $\theta$-rotation is defined in equation \ref{ellipsoid}), we observe a delay in the initiation of the impact forces for Spheroid-45 and Spheroid-75, though negligible on the latter case. Overlooking this delay in time, the trend observed in the previous figure \ref{force_PS} still applies here, namely, for the cases which reach higher peak forces earlier, the subsequent decay stage after the peak force is faster. This is again due to the conserved momentum change during the impact event. Calculations on the integral of the impact force over time provide $2.09$ for Spheroid-45 and $2.10$ for Spheroid-75, both of which match well with the theoretical value in equation \ref{mom}.

\subsubsection{Effect of droplet volume}\label{3.2.3}
When comparing cases $1$ and $7$, the droplet sizes are scaled with a ratio of $\lambda=2/\sqrt[3]{2}$ while other parameters remain unchanged. This indicates their dimensionless flow parameters, including $Re$, $We$, and $Fr$, are within the same order of magnitude. Therefore, we expect the flow behaviours to be qualitatively identical. Particularly, for physical quantities that relate to impact loadings on the solid surface, after particular scalings of their radii, should quantitatively collapse onto each other.

Figure \ref{force_PV} compares the impact force on the solid surface between the Sphere and Sphere-s cases. After scaling the dimensionless impact force, $f$, by radius squared, $r_xr_z$, and dimensionless time, $t$, by the vertical radius, $r_z$, two temporal force evolutions collapse onto the same curve, as expected.

Having verified the existence of the scaling law between Sphere and Sphere-s, we do not compare further between the two cases. However, we note that, in reality, the difference between droplets impacts of different sizes may not trivially obey the scaling law due to the interplay of additional factors, \textit{e.g.}, cushion effects of the entrapped air or liquid film  \cite{Nils2023}.

\section{Discussion}\label{discussion}
\subsection{Universal scaling law}

\begin{figure}
\centering
\includegraphics[width=9.5 cm]{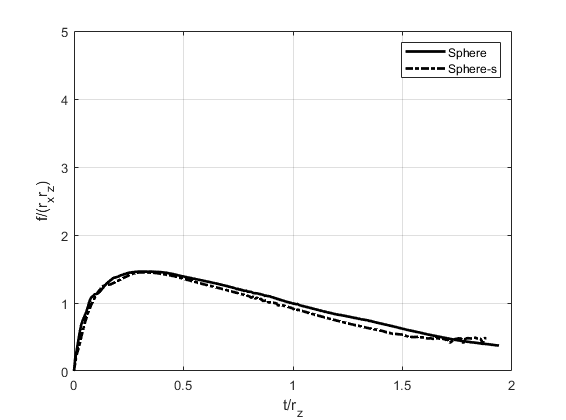}
\caption{The scaled impact force evolutions, $f/(r_x r_z)$, in the scaled time, $t/r_z$, in the two parametric cases: Sphere and Sphere-s }\label{force_PV}
\end{figure}

\begin{table*}
\caption{Summary of the effective radii $e_x$ and $e_y$, projection area, influential radii $i_x$ and $i_z$, (approximate) influential area, pressure peaks $P_{\text{max},x}$ and $P_{\text{max},z}$ in $x$- and $y$-directions respectively, and the overall maximum pressure in the parametric cases. The (approximate) influential area is calculated by $\pi i_x i_z$, and the overall maximum pressure is the bigger value of $P_{\text{max},x}$ and $P_{\text{max},z}$.}
\centering
\begin{tabular}{ccccccc} 
\toprule
 & \textbf{Case $1$} & \textbf{Case $2$} & \textbf{Case $3$} &  \textbf{Case $4$} &  \textbf{Case $5$} &  \textbf{Case $6$}\\
& \textbf{Sphere} & \textbf{Ellipsoid} & \textbf{Spheroid-0} &  \textbf{Spheroid-45} &  \textbf{Spheroid-75} &  \textbf{Spheroid-90}\\
\midrule
\textbf{$e_x$} &  $1/\sqrt[3]{2}$ &$1/\sqrt{2}$&$1$&$\approx0.79$&$\approx0.55$&$1/2$\\
\midrule
\textbf{$e_z$} &  $1/\sqrt[3]{2}$ &$1$&$1$&$1$&$1$&$1$\\
\midrule
\textbf{Projection Area, $\bm{\pi e_x e_z}$} &  $1.98$ &$2.22$&$3.14$&$2.48$&$1.73$&$1.57$\\
\midrule
\textbf{$i_x$} &  $0.32$ &$0.27$&$0.43$&$0.25$&$0.21$&$0.20$\\
\midrule
\textbf{$i_z$} &  $0.32$ &$0.39$&$0.43$&$0.48$&0.27&$0.41$\\
\midrule
\textbf{Influential Area, $\bm{\pi i_x i_z}$} &  $0.31$ &$0.33$&$0.59$&$0.38$&$0.18$&$0.27$\\
\midrule
\textbf{$P_{\text{max},x}$} &  $4.49$ &$4.90$&$7.68$&$5.31$&$3.39$&$3.13$\\
\midrule
\textbf{$P_{\text{max},z}$} &  $4.49$ &$4.91$&$7.68$&$2.98$&$2.72$&$3.18$\\
\midrule
\textbf{Overall Maximum Pressure} &  $4.49$ &$4.91$&$7.68$&$5.31$&$3.39$&$3.18$\\
\bottomrule
\label{table_pro}
\end{tabular}
\end{table*}

In section \ref{results}, we found that the spatio-temporal evolutions of impact pressure and force on the solid surface vary with the droplet shape and impact orientation (\textit{i.e.}, rotation angle) in general. In particular, we noticed the trend that, for droplets reaching higher impact pressure (or force) peaks, the instants of the pressure (or force) peaks happen earlier in time. Apart from that, we can also see from figures \ref{figure_PS} and \ref{figure_PA} that the influential areas, defined as the area of the solid surface with impact loadings, at the instants of pressure peaks positively correlate with their pressure peak magnitudes. This reminds us the scaling law observed in section \ref{3.2.3}, which interprets a positive relation between droplet size (and hence influential area) and impact loadings, including pressure, force and momentum. Nevertheless, this scaling law was obtained on the assumption of identical droplet shapes. In this section, we seek to explain the observed two trends from a physical perspective and, on this basis, further explore the existence of a \textit{universal scaling law} between droplets of distinctive shapes.

We start from the cases of Spheroid-0, Spheroid-45, Spheroid-75, and Spheroid-90, where the droplet shapes are the same but rotation angles are different. Their \textit{effective radii}, $e_x$ and $e_z$, defined as the radii of the droplets' projections (before impact) on the solid surface, are listed in table \ref{table_pro}, together with their projection areas, $\pi e_x e_z$. Another reason to start with these four cases is that their effective radii in $z-$direction are identical, so the trend of effective radii in $x-$direction is simply that of the projection areas. Table \ref{table_pro} also lists their \textit{influential radii}, $i_x$ and $i_z$, defined as a half of the distance between the two roots of the peak pressure distributions in $x-$ and $y-$directions, respectively, in figures \ref{figure_PS} and \ref{figure_PA}. The product of influential radii, $\pi i_x i_z$, provides an approximation for the influential area. As we can see, from case $3$ to case $6$, the influential radii in $x-$direction, $i_x$, monotonically decrease with the effective radii in $x-$direction, $e_x$. While in the $z-$direction, for identical $e_z$ in all four cases, only $i_z$ from Spheroid-0 and Spheroid-90 are similar. This is because the asymmetric cases of Spheroid-45 and Spheroid-75 have the lowest points of geometries, which hit the solid surface first, off the line $x=0$ (see red dashed lines in the insets of figure \ref{figure_PA}), and are hence not comparable.

For more complex cases of Sphere, Ellipsoid, Spheroid-0 and Spheroid-90 where mixed trends happen in $x-$ and $z-$directions, the same rule applies. In the $x-$direction, Spheroid-0 has the largest $e_x$ and $i_x$, followed by Sphere, Ellipsoid, and Spheroid-90 in the order of decreasing magnitudes. While in the $y-$direction, Ellipsoid, Spheroid-0, and Spheroid-90 (with equally larger $e_z$ than that of Sphere) have similar and predominantly larger $i_z$ than that of Sphere. As a result, the influential areas of the four cases are in line with the order of their projection areas.

\begin{figure}
\centering
\includegraphics[width= 9.5cm]{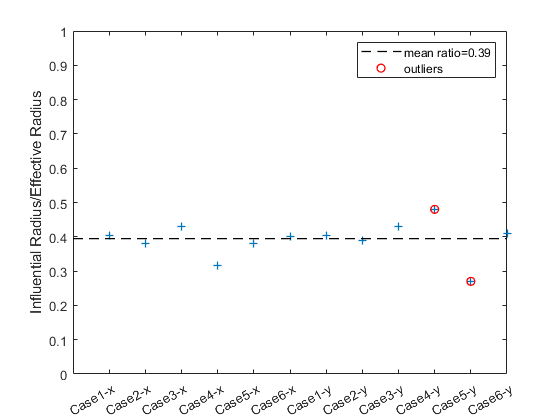}
\caption{The ratios between influential radius and effective radius in both $x$- and $y$-directions among all the parametric cases. The mean value of the ratios is $0.39$ with a CV of $8\%$.}
\label{figure_scale}
\end{figure}

To quantify this universal observation, we divide the influential radius by the effective radius in both $x-$ and $z-$-directions for all cases, and present the obtained ratios in figure \ref{figure_scale}. As can be seen, apart from the two outliers\textemdash the $z-$direction of Spheroid-45 and Spheroid-75, whose deviations are explained above (figure \ref{figure_PA})\textemdash all cases have the ratios in both directions collapse onto a nearly constant value, yielding a mean of $\alpha=0.39$ and a coefficient of variation (CV) of only $8\%$ (excluding two outliers), defined as the standard deviation divided by the mean.

\begin{figure}
\centering
\includegraphics[width=9.5 cm]{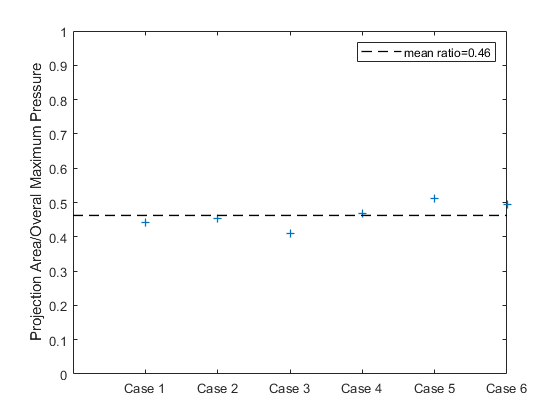}
\caption{The ratios between projection area and overall maximum pressure among all the parametric cases. The mean value of the ratios is $0.46$ with a CV of $8\%$.}
\label{figure_scale2}
\end{figure}

Regarding the impact pressure peaks in the two directions, we noticed that they are more of an overall facial behaviour on the solid surface. That said, we only compare the overall maximum pressure (see the last row of table \ref{table_pro}) on the two lines, $x=0$ and $z=0$, \textit{i.e.}, the bigger value of $P_{\text{max},x}$ and $P_{\text{max},z}$, which reasonably characterises the maximum impact pressure on the entire solid surface, with the effective projection area for all cases. Another reason to focus on the overall maximum pressure is because of its particular relevance to impact erosion studies, where the most severe loading is often of primary concern. Indeed, by doing so, we found that the sequence of the overall maximum pressure is consistent with the sequence of projection area: (in decreasing magnitudes) Spheroid-0, Spheroid-45, Ellipsoid, Sphere, Spheroid-75, and Spheroid-90. Moreover, quantitatively, as shown in figure \ref{figure_scale2}, their ratios between the projection area and overall maximum pressure nicely collapse onto a nearly constant value, with a mean of $\beta=0.46$ and a CV of only $8\%$, again.

We note that the findings on the ratios in figures \ref{figure_scale} and \ref{figure_scale2} are regardless of the droplet shape and the orientation angle, at least for the cases investigated in the present study. Thereby, combining two ratios, we derive a universal formula between the overall maximum pressure and the influential radii at corresponding peak force instants of each direction, $i_x$ and $i_z$, for arbitrary droplets, as:
\begin{equation}
	Overal\, Maximum\, Pressure = \frac{\pi (i_x/\alpha) (i_z/\alpha)}{\beta} =  \frac{\pi}{\alpha^2\beta}i_x i_z
\label{equ_law}
\end{equation}
with $\alpha=0.39$ and $\beta=0.46$ from fittings. Equation \ref{equ_law} can be used to approximate the maximum impact pressure on the solid surface for non-axisymmetric cases. The approximation is expected to be more accurate if the lowest point of the droplet geometry is on the $y-$axis, and almost exact if the droplet is axi-symmetric.

\begin{figure}
\centering
\includegraphics[width=9.5 cm]{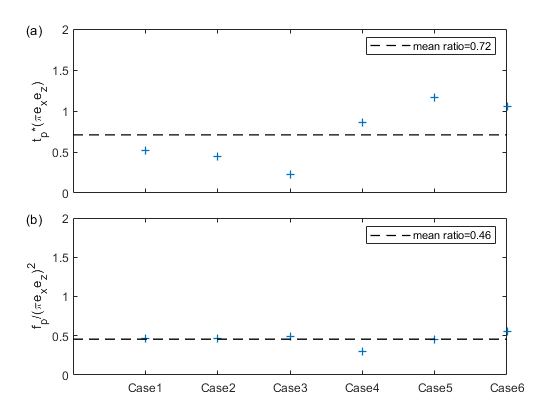}
\caption{($a$) The ratios between peak force instant, $t_p$, and the inverse of projection area, $1/(\pi e_x e_z)$ among all the parametric cases. The mean value of the ratios is $0.72$ with a CV of $48\%$. ($b$) The ratios between peak force, $f_p$, and the square of projection area, $(\pi e_x e_z)^2$ among all the parametric cases. The mean value of the ratios is $0.46$ with a CV of $19\%$.}
\label{figure_scale3}
\end{figure}

\subsection{Impact force}
We reported a positive correlation between the maximum droplet impact pressure on the surface and its effective radii of the initial condition. Nevertheless, since the total liquid volume is conserved for each droplet, this higher intensity of impact loadings must be compromised in other ways. As a result, we noticed in figures \ref{force_PS} and \ref{force_PA} that the peak force instants, $t_p$, which effectively characterise the lengths of impact durations on the surface happen earlier in time for the more intensified cases with higher impact force peaks. In this section, we therefore aim to quantify this \textit{accompanying effect} on impact force associated with the universal scaling law, including the peak force instant $t_p$ and magnitude $f_p$.

Impact force is the surface integral of the impact pressure on the solid surface. Hence, by the universal scaling law, namely the impact pressure on the surface is proportional to the projection area, we expect the impact force to be proportional to the square of the projection area: 
\begin{equation}
    f\propto (\pi e_x e_z)^2
    \label{RF}
\end{equation}
While impact time, as shown in figure \ref{force_PV},  scales with the vertical radius, and is hence inversely proportional to the projection area:
\begin{equation}
    t \propto e_y \propto \frac{1}{e_x e_z} \propto \frac{1}{\pi e_x e_z}
    \label{RT}
\end{equation}
The second proportionality in equation \ref{RT} holds exactly for axisymmetric ellipsoids of the general form shown in figure \ref{parametric}, where the effective radii are identical to the actual radii, since the total volume, $v_0=(4\pi/3) r_x r_y r_z=(4\pi/3) e_x e_y e_z$, is conserved. For other geometries, equation \ref{RT} remains a good approximation, as we verify \textit{a posteriori}.

Figure \ref{figure_scale3}($a$) presents the ratio between the peak force instant, $t_p$, and the reciprocal of the projection area, $1/(\pi e_x e_z)$, for all six cases of different droplet shapes. The resulting ratios collapse satisfactorily onto a nearly constant value, with a mean of $0.72$ and an acceptable CV of $48\%$. Although this relatively large CV suggests that the underlying relationship may deviate from the simplified inverse-proportional scaling, the proposed scaling remains a useful and straightforward approximation. Figure \ref{figure_scale3}($b$) presents the ratio between the impact force peak, $f_p$, and the square of the projection area, $(\pi e_x e_z)^2$, for all the cases. Again, the resulting ratios nicely collapse onto a nearly constant value, with a mean of $\gamma=0.46$ and a CV of $19\%$. The good collapses in these two plots verified the scaling laws in equations \ref{RF} and \ref{RT}, including the non-axisymmetric cases of Spheroid-45 and Spheroid-75. 

\begin{figure}
\centering
\includegraphics[width=9.5 cm]{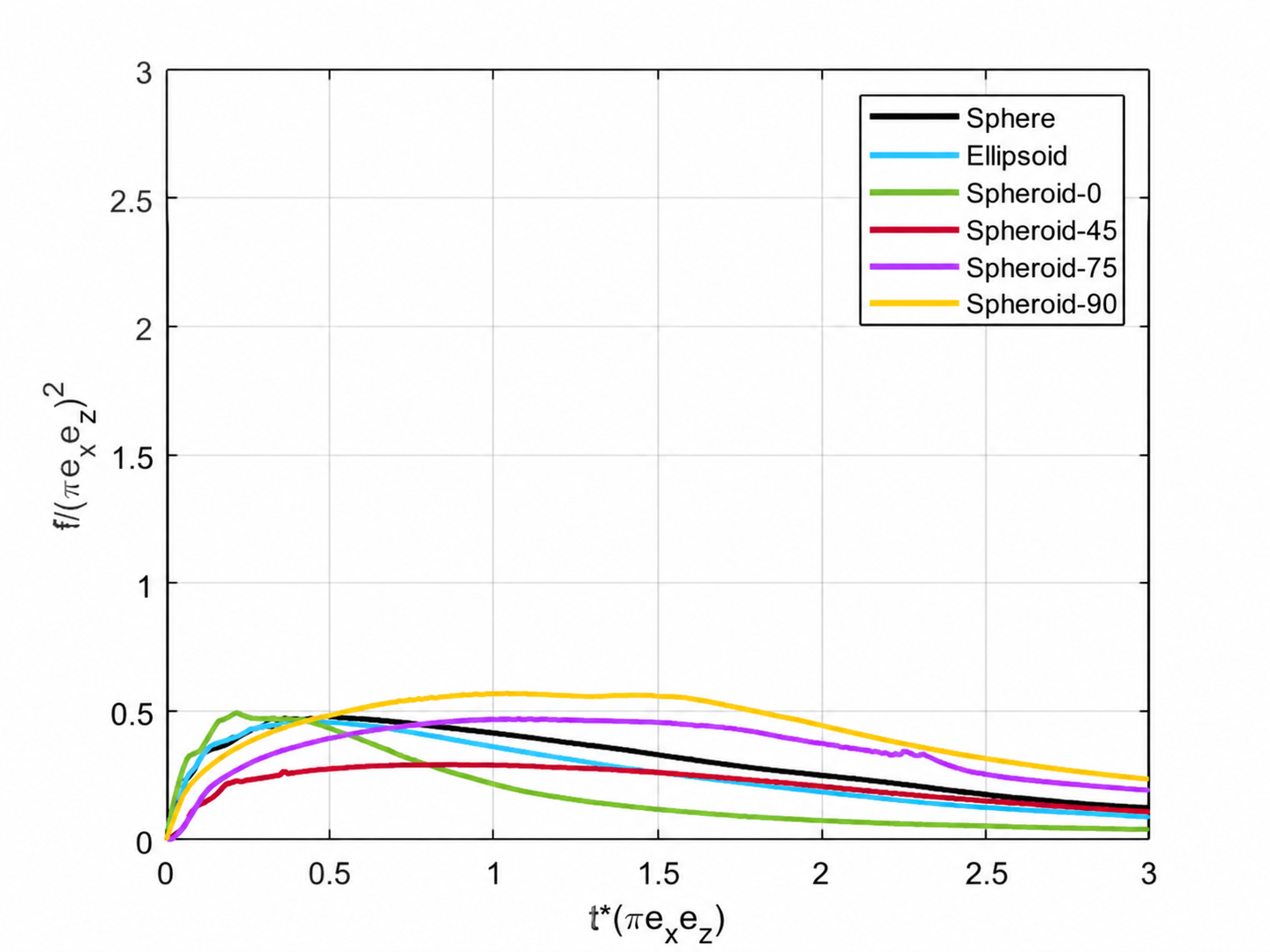}
\caption{The scaled impact force temporal evolution, $f/(\pi e_x e_z)$, with the scaled time, $t(\pi e_x e_z)$, for all parametric cases. The force temporal evolutions collapse onto a nearly constant scaled peak of value $0.46$.}\label{force_scale}
\end{figure}

Combining the two scaling laws in equations \ref{RF} and \ref{RT}, figure \ref{force_scale} shows the scaled impact force evolutions, $f/(\pi e_x e_z)^2$, plotted against the scaled impact time, $t(\pi e_x e_z)$. Except for the two non-axisymmetric cases (Spheroid-45 and Spheroid-75), all cases collapse well onto a common curve during the early stage of impact, up to the scaled peak force at $\gamma=0.46$, although their subsequent decay tails diverge at different rates. This divergence is primarily attribute to the relatively large CV associated with the fitting of equation \ref{RF}, which suggests that the peak force instant deviates to some extent from the simplified inverse-proportional scaling. Nevertheless, the overall agreement in figure \ref{force_scale} indicates that the proposed scaling laws capture the temporal evolution of the impact force reasonably well up to the peak instant, a period that is particularly relevant for impact erosion studies.

To summarise, the findings demonstrate that key impact loading characteristics can be related to the droplet projection area, which can be estimated from the effective radii determined by the initial droplet geometry and impact orientation. The proposed scaling laws therefore provide a predictive framework for estimating the peak impact pressure, peak impact force, and characteristic impact duration directly from the initial droplet conditions, without requiring a separate high-resolution simulation for each configuration. This capability is particularly valuable in practical impact scenarios, where droplets may deviate substantially from spherical shapes and strike surfaces in different orientations.

From an erosion assessment perspective, these loading characteristics can be incorporated into material-specific erosion criteria or damage models to evaluate the susceptibility of a surface to impact-induced damage. For example, the predicted pressure and force histories can serve as inputs to fatigue lifetime models to estimate material lifetimes under impact conditions more representative of real-world engineering applications. The proposed framework thus establishes a link between droplet impact dynamics and erosion assessment, providing a basis for more efficient evaluation of impact-induced surface damage.

\section{Conclusion}
\label{conclusion}
In this paper, we developed a fluid-structure interaction (FSI) model to study liquid droplet impacts on solid materials. Using this model, we conducted a parametric study of droplets with varying shapes and orientation angles. The aim was to identify a universal scaling law that links the spatio-temporal evolution of impact loadings on the solid surface to the initial droplet geometry.

We found that, under volume conservation, changes in droplet shape lead to a positive correlation between the peak impact pressure on the surface, the influential area at the instant of pressure peak, and the peak impact force. In contrast, the time of peak force scales inversely due to momentum conservation. 

To rationalise these trends, we proposed a \textit{universal scaling law} that relates the spatio-temporal evolution of impact loadings\textemdash including pressure and force distributions\textemdash to the \textit{projection area} of the droplet. Specifically, the influential area at the instant of pressure peak is proportional to the projection area and to the surface pressure peak; the impact force peak scales with the square of the projection area; and the time of peak force is inversely proportional to the projection area. We carefully determined the corresponding proportionality constants by fitting.

The findings show that all key impact loadings can be expressed in terms of the projection area, which in turn can be estimated from the \textit{effective radii} determined by the initial droplet geometry and impact orientation. Thus, the proposed universal scaling law provides a predictive framework for impact loading dynamics based solely on initial droplet conditions of arbitrary shapes. This approach is expected to be particularly valuable for improving the accuracy of impact erosion studies, where liquid droplets often strike with non-ideal shapes.

\section*{Funding}
H.H. was funded by the PhD Scholarship at the Department of Mechanical Engineering of Imperial College London.



\end{document}